\documentclass[letterpaper]{article}
\usepackage{aaai18} 
\usepackage{times}  
\usepackage{helvet} 
\usepackage{courier} 
\usepackage{url}  
\usepackage{graphicx}  
\usepackage{paralist}
\usepackage{booktabs} 
\usepackage{dblfloatfix}
\usepackage{graphicx}
\usepackage{amssymb}
\usepackage{array}
\usepackage{amsmath}
\usepackage{filecontents,lipsum}
\usepackage{amsmath}
\usepackage[linesnumbered,ruled]{algorithm2e}
\title{Intelligently Assisting Human-Guided Quadcopter Photography}
\author{Saif Alabachi\\ University of Central Florida, Orlando\\ University of Technology Baghdad\\ \url{s.mohammed@knights.ucf.edu} \And Gita Sukthankar\\ University of Central Florida\\ Orlando, FL\\ \url{gitars@eecs.ucf.edu}}

\begin{document}
\maketitle
\begin{abstract}
Drones are a versatile platform for both amateur and professional photographers, enabling them to capture photos that are impossible to shoot with ground-based cameras.   However, when guided by inexperienced pilots, they have a high incidence of collisions, crashes, and poorly framed photographs.  This paper presents an intelligent user interface for photographing objects that is robust against navigation errors and reliably collects high quality photographs.  By retaining the human in the loop, our system is faster and more selective than purely autonomous UAVs that employ simple coverage algorithms.  The intelligent user interface operates in multiple modes, allowing the user to either directly control the quadcopter or fly in a semi-autonomous mode around a target object in the environment.  To evaluate the interface, users completed a data set collection task in which they were asked to photograph objects from multiple views.  Our sketch-based control paradigm facilitated task completion, reduced crashes, and was favorably reviewed by the participants.
\end{abstract}

\maketitle
\section{Introduction}
\begin{figure}
\centering
\includegraphics[width=7cm,height=9cm]{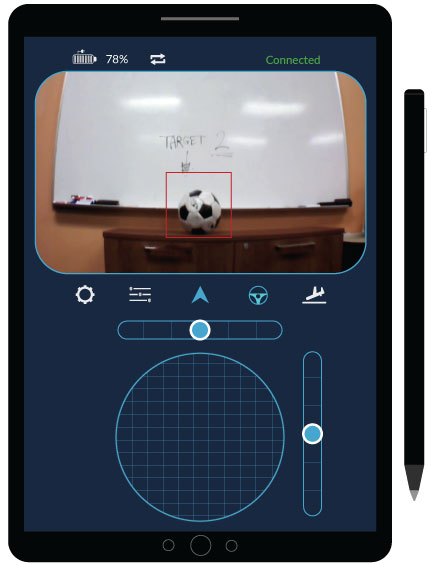}
\caption{The upper part of the interface broadcasts the image stream from the quadcopter frontal camera. The user can select a target by drawing a bounding box over the camera view.  At any time, the human can directly control the quadcopter by sketching on the lower control panel.}
\label{fig:sui}
\end{figure}

Under the supervision of a careful, experienced pilot, quadcopters can be used to capture amazing photographs; however, the typical user's experience is marred by crashes and poor quality photos.\footnote{See the NYTimes article ``Santa Delivered the Drone. But Not the Safety and Skill to Fly Them.''.}  The question remains----how to provide a rewarding human-robot interaction for inexperienced users working with hobbyist quadcopters?   This paper proposes a sketch-based interface which is designed to hide certain degrees of freedom from user control and prevent crashes by monitoring the scale of the targeted object (Figure~\ref{fig:sui}).  

The interface offers the following functionality: 1) three canvases for manual navigation that capture user sketches and translate them into control commands in eight directions; 2) a canvas for displaying the real-time image from the quadcopter frontal camera that can be used to select an object of interest for the quadcopter to track autonomously; 3) a dataset collection mode in which the quadcopter autonomously collects images suitable for a variety of applications, including infrastructure inspection, 3D reconstruction, and training machine learning classifiers.  The system is implemented as a web-based application that can be run on a variety of platforms with no installation required.  It can be accessed by multiple clients, allowing several users to cooperatively direct the quadcopter.  Rapid and reliable object tracking is achieved through the use of an adaptive correlation filter (MOSSE, Minimum Output Sum of Squared Error~\cite{bolme2010visual}); previous systems have relied on color-based tracking strategies~\cite{kim2013vision}.  This paper compares the performance of our interface vs.\ two commercial drone control systems.  Based on the participants' performance on indoor image collection tasks, we believe that our system improves on commercial options and provides precise control at low cost without additional hardware extensions.

\section{Related Work}

The problem of creating autonomous robot photographers both mobile~\cite{byers2003autonomous,campbell2005} and aerial~\cite{srikanth2014computational,Sukthankar-Rey-FLAIRS2016}) has been examined by several researchers.  In some application domains, it is feasible for a quadcopter to fly completely autonomously, particularly when performing high altitude visual surveillance or mapping~\cite{huang2011visual} tasks.  One specific problem, tracking and photographing humans, is particularly interesting since humans are good subjects for photography and can be easier to track.

However, there are a variety of visual inspection and surveying tasks that require photographing arbitrary objects; our interface is specialized for handling those types of problems.   We considered many candidate interface modalities when designing our system, including gesture, voice, gaze and EEG, which have been successfully employed in other quadcopter systems.  Unlike many sketch-based robot control systems (e.g., \cite{sakamoto2009sketch,cummings2012sketch,richards2015user})) in our system the user designates the target on a canvas displaying the robot's view rather than on the global map.  Our system is most similar to XPose, a touch-based system for interactive photo taking~\cite{xpose}; however unlike XPose, our user interface does not require global localization as the selected object is used as a position reference.
  
\section{Method}

\begin{figure*}
\centering
	\includegraphics[width=0.88\textwidth]{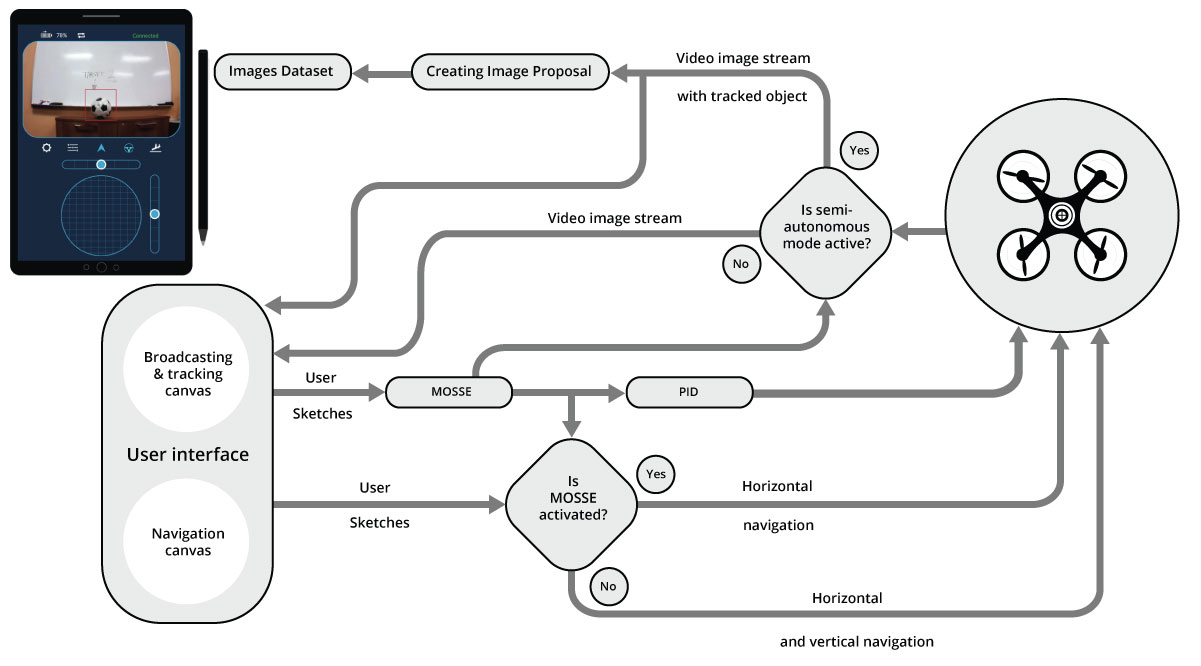}
\caption{The user interface accepts two categories of user sketches: 1) navigation strokes from the three designated canvases which specify the direction and velocity of the quadcopter and 2) the boundary strokes on the broadcasting canvas that enclose the area of interest. During semi-autonomous operation, the MOSSE adaptive correlation filter outputs the object centroid point and the corresponding bounding box. The broadcasting canvas receives the raw image with embedded tracking results at each time interval $t$. If the object is tracked successfully, the navigation agent locks the yaw angle and altitude of the quadcopter and calculates the measured centroid error to transmit to the PD controller.}
\label{fig:platform}
\end{figure*}

Figure~\ref{fig:platform} shows the system architecture.  Our experiments were performed on the commercially available Parrot Augmented Reality (AR) Drone Version 2.
This drone has two cameras: one front-mounted HD camera and a downward facing QVGA camera. The on board battery provides 15 minutes of continuous flight.  The Parrot AR Drone has a 1 GHz ARM Cortex A8 processor and 1 Gbit DDR2 RAM; it runs GNU/Linux and connects to a laptop over the wireless LAN (see \citeauthor{piskorski2012ar} \citeyear{piskorski2012ar} for the complete list of hardware and software specifications).  The back end was constructed on top of the Robot Operating System (ROS) which can handle communication between several entities without experiencing significant latency.  We created a web server using ROS Web Tools~\cite{alexander2012robot} and assigned a specific HTTP port to emit ROS video streaming messages, while using network web socket tools to communicate to our user interface.

\subsection{Sketch-based User Interface}
For the front end of our system, we designed a web-based interface, which can be simultaneously accessed by multiple users, using a variety of mobile devices.  The ARDrone utilizes User Datagram Protocol (UDP) to communicate with the ground analysis and command unit (back-end), and HTTP is used for data transmission between the front-end and the ground (back-end) unit.  There is a dedicated canvas showing the view from the quadcopter front camera, along with three areas that control translation, yaw, and altitude.  To enter autonomous object tracking mode, the user can circle a region in the front view canvas.

The user can assert direct control by sketching in the lower panel; there are separate controls for stop, go, takeoff, and landing.  The sketches are then translated into linear and angular velocities in the quadcopter coordinate system and normalized by the total length of the corresponding canvas.


\subsection{Autonomous Object Photography Mode}
For the vision system, we evaluated several object detection and tracking approaches before deciding to use an adaptive correlation filter (MOSSE) to track the region enclosed by the user on the front-view canvas.  MOSSE~\cite{bolme2010visual} employs convolution to perform the tracking, after creating an appearance model with adaptive correlation filters.  The simplicity of the procedure allows MOSSE to track objects in video captured at high frame rates ($>600$ frames per second (fps)).   The appearance model is trained in the Fourier domain using a set of random affine transformations, and the aim is to minimize the sum squared error between the desired and actual convolution outputs.  During the tracking process, three ROS messages are created for each $t$ period: 1) the centroid point of the tracked object, 2) a tuple-type message for streaming the bounding box coordinates, and 3) an image-type message containing both the front camera image and the bounding box to be viewed on the user interface.  $x_{\text{min}}$, $y_{\text{min}}$, $x_{\text{max}}$, and $y_{\text{max}}$ are extracted from the circle stroke, and that region of the image is used to initialize the adaptive correlation filter.

After the initial bounding box is drawn, the quadcopter starts flying autonomously, and the system enters a visual dataset collection mode, acquiring data at a rate of 1 fps. The quadcopter modifies its yaw angle and altitude to track the object designated by the user.  The x-axis error between the object centroid and canvas center is used to estimate the orientation angle, and the y-axis error is used to estimate the quadcopter's altitude.  
\begin{equation}
	\text{error}_{x} =  (x_{\text{centroid}} - x_{\text{center}}) / x_{\text{max}}
\end{equation} 
\begin{equation}
	\text{error}_{y} =  (y_{\text{centroid}} - y_{\text{center}}) / y_{\text{max}}
\end{equation} 
The errors are transmitted to a PD (proportional-derivative) controller with gains $K_p$ and $K_d$ set to 0.25.  The quadcopter uses its inertial sensors to monitor roll $\Phi$, pitch $\Theta$, yaw $\psi$, rotational speed $\Psi$ and the vertical velocity $\zeta$; controls are issued using a series of ROS Twist commands $u = (\bar{\Phi}, \bar{\Theta}, \bar{\zeta}, \bar{\Psi}) \in [-1,1]^4$ at a frequency of 100Hz.  Our interface is capable of eliminating undesired photos by comparing the correlation percentage to a predefined threshold; as long as this percentage exceeds the specified threshold, the agent continues photographing the tracked object, else it stops.

\section{Evaluation}  

\begin{figure}
\centering
\includegraphics[width=8.25cm,height=5.5cm]{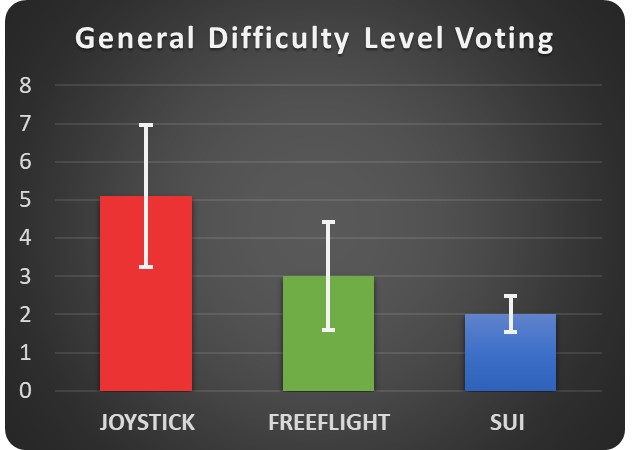}
\caption{Mean and standard deviation obtained from ten participants' rating of the difficulty of each control modality (Study 2 and 3), where 7=most difficult to use and 1=easiest to use.  Our interface (SUI) was rated by the users as being significantly easier to use according to a single tailed paired t-test ($p<0.05$).}
\label{fig:q1}
\end{figure}

We sought feedback on our user interface design from three groups of users.   First, to evaluate the ease of learning our sketch-based control paradigm, an observation study was conducted on a group of elementary/high school lab visitors who were asked to fly the quadcopter to a target and land it.   In the second study, the performance of the sketch-based user interface was compared to the performance of joystick control for piloting the drone.  In the third study, the autonomous visual data collection was evaluated vs.\ AR.Free Flight image capture.  All experiments were performed in an indoor environment, and users were trained in the usage of each control paradigm for five minutes before commencement of testing.  Pre and post questionnaires were administered during the second and third studies.  Figure~\ref{fig:q1} shows the participants' ratings of the difficulty of aerial control under each control modality (joystick, AR.Free Flight, and our smart user interface (SUI)); our interface was rated by ten users as being significantly easier to use ($p<0.05$ on a single tailed paired t-test).  A video demo of our system can be viewed at: \url{https://youtu.be/ErA2111xjzMl}.

\begin{table}
\centering
\begin{tabular}{ |p{2.5cm}|p{1.5cm}|p{2.5cm}|  }
\hline
\multicolumn{3}{|c|}{Study 1} \\
\hline
Commands    		 & Stroke & Bounding Box\\
\hline
Take Off	 		 & 6/6 & 6/6\\ 
Navigation       		 & 4/6 & 4/6\\
Reach the Goal   		 & 3/6 & 4/6\\
Landing      	 & 2/6 & 5/6\\
Interest         	 	 & 6/6 & 5/6\\ 
\hline
\end{tabular}
\caption{Performance of elementary/high school children using the sketch-based interface. Six children participated, and we tallied how many of the quadcopter control tasks they were able to perform, as well as their interest in the system.  In the first condition they were asked to just use the stroke control.  Then they were allowed to use a simple bounding box control system, similar in concept to the vision-based tracking but without the filtering or the image capture.   Since the children performed well on most of the elements using the bounding box, we decided to incorporate it into our final design.}
\label{table:1}
\end{table}

\subsection{Study 1: Elementary/High School Observation} 
Our elementary/high school guests included four males and two females between the ages of 12 and 16 years old. Our main goal for this study was to observe how younger users would perform with the sketch-based control.  The participants were given five minutes of practice and then were asked to try two flying procedures.  The first procedure was to fly the quadcopter in a circuit by sketching strokes on the navigation canvases.  In the second procedure, they were asked to select a target by sketching a bounding box, as part of flying the circuit. The quadcopter then tries to center the target using a simple visual servoing algorithm. 

Table~\ref{table:1} shows the participants' performance in achieving the required tasks for both procedures. The AR.Drone 2.0 elite edition has a maximum speed of about eleven meters per second, which is quite high when navigating in an indoor environment. For more safety, we added an option to our system in which the user can limit the speed by curtailing the length of drawn strokes.   This study enabled us to test whether using the bounding box to guide the quadcopter was an intuitive control choice.   We determined that the addition of that control option improved navigation, particularly for promoting successful landings.

\subsection{Study 2: Navigation Control}

\begin{figure}
\centering
\includegraphics[width=8.8cm,height=9cm]{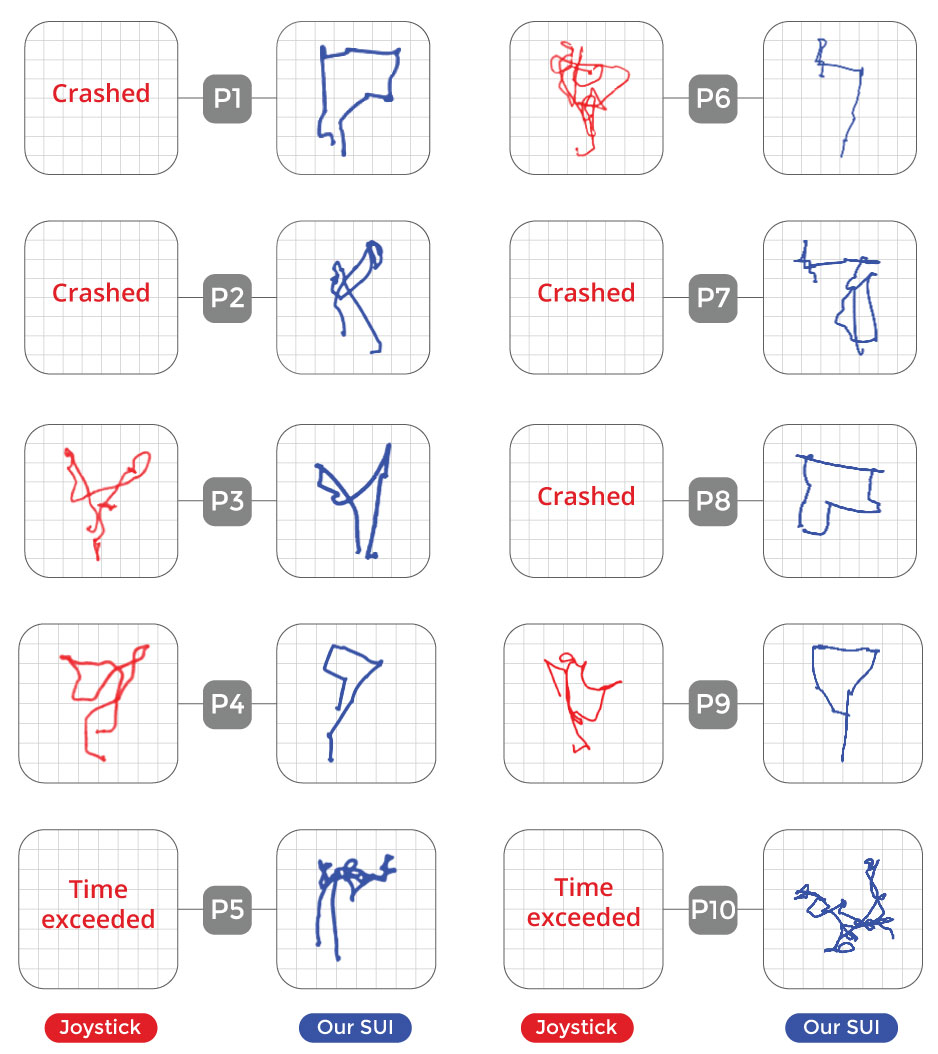}
\caption{In study 2, participants were asked to fly to two target objectives and return to the start point using joystick control and our sketch-based user interface.  We exported the flight paths that the quadcopter measured using its SLAM system.  An ideal path would be shaped like an isosceles triangle. The red paths are the ones executed under joystick control, and the blue ones were done with our user interface. Paths from participants P1, P2, P5, P7, P8 and P10 were not captured because either they were unable to reach the targets using a joystick within the specified time or crashed the drone three times.}
\label{fig:s2}
\end{figure}

Our second study focused on evaluating navigation performance with the user interface. Our participants (Table~\ref{table:2}) were assigned two objectives to reach with the quadcopter.  The first object was a fire-alarm mounted on the wall, and the second one was a soccer ball placed on a cabinet.  They were asked to fly the AR.Drone, face each target while maintaining a safe one meter distance, and return to the start point.  For evaluation purposes, we employed a SLAM system~\cite{klein2007parallel} to estimate the position of the quadcopter during flight (Figure~\ref{fig:s2}).

\begin{table}
\centering
\includegraphics[width=8cm,height=4cm]{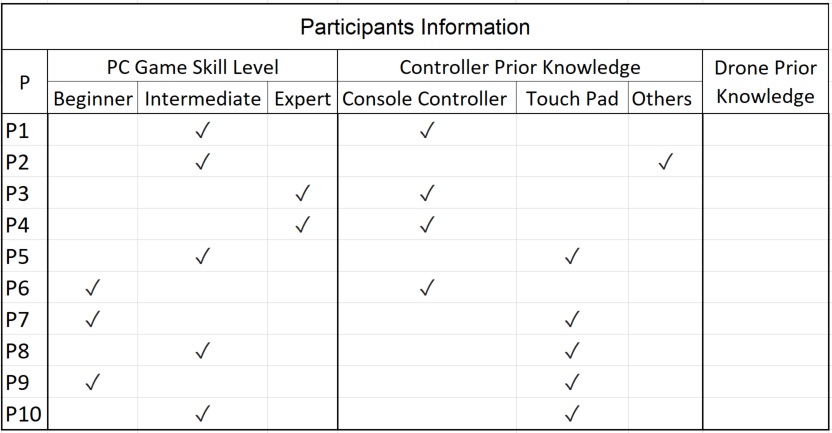}
\caption{Participant information for the second and third study.  None of the users had prior experience flying drones.  Many of them were intermediate or expert game players.  All participants but one felt most comfortable with either console or touch pad game controllers.}
\label{table:2}
\end{table}

\begin{figure}
	\centering
	\includegraphics[width=8cm,height=5cm]{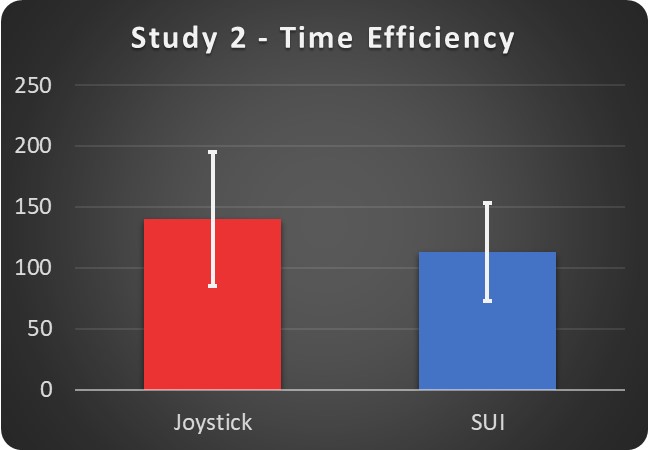}
	\caption{Average time required (seconds) for participants to complete both navigation tasks in study 2.  Participants that crashed the quadcopter were assumed to have taken the maximum required time (180 seconds).   Despite a few fast joystick runs by the expert users, our user interface led to a faster average completion time ($p<0.05$ according to a paired single-tailed t-test)}.
	\label{fig:q2}
\end{figure}

Users flew the scenario once using the joystick control and the other time using our interface (in randomized order).  Two participants rated themselves as expert gamers in the pre-questionnaire.  They were able to fly acceptably well using the joystick, but many of the other users either crashed or exceeded the allotted time. However, with our interface, all participants were able to fly the quadcopter without crashing and complete the task in under the three minute time limit (Figure~\ref{fig:q2}).  The PR (percentage of targets reached) was measured, along the time required, including overtime trials and crashes (Table~\ref{tab:t3}).  Participants using the interface had higher success rates at reaching the targets, compared to joystick control. Expert users experienced slightly slower flight times, however when accounting for unsuccessful trials the overall time required for our user interface was improved.

\begin{table}
\centering
\begin{tabular}{ |c|c|}
			\hline
			\multicolumn{2}{|c|}{Study 2 (Navigation Performance)} \\
			\hline
User Interface  & Targets Reached (\%)\\ 
\hline
Joystick		 	 & 40\%\ \\
Interface       		 & 100\%\ \\
\hline
\end{tabular}
\caption{Our user interface makes navigation much more reliable for the users.  In Study 2, only 40\% of the targets were reached (across all users), whereas 100\% of the targets were reached by participants employing our user interface.}
\label{tab:t3}
\end{table}

\begin{figure}
\centering
\includegraphics[width=8cm,height=5cm]{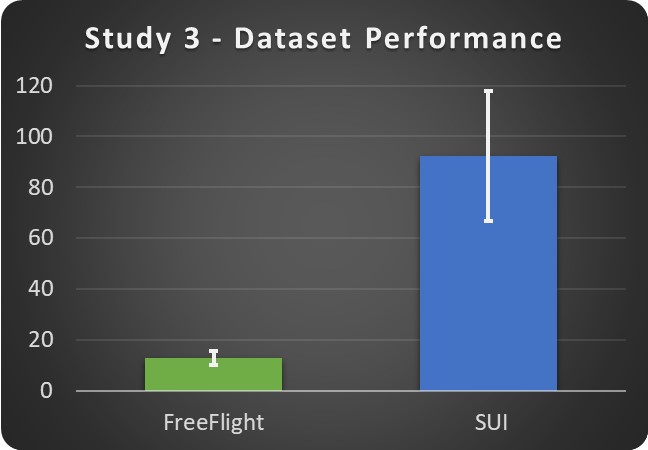}
\caption{Number of images collected by users in study 3.  Our interface supports more prolific image collections which are useful for training machine learning classifiers.}
\label{fig:q3}
\end{figure}

     \begin{figure*}
	\centering
	\includegraphics[width=18cm,height=6cm]{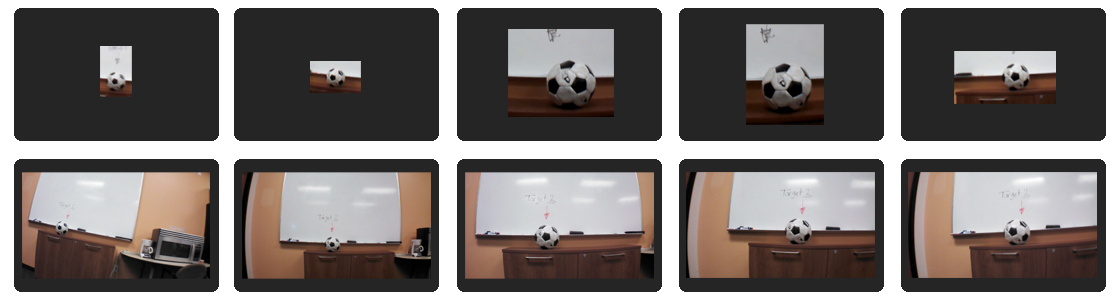}
\caption{Examples of images captured using the visual dataset collection mode.  The top row shows images captured with our user interface; the bottom row shows images captured with AR.FreeFlight.}
	\label{fig:s3}
\end{figure*}

\begin{figure}
	\centering
	\includegraphics[width=8cm,height=5cm]{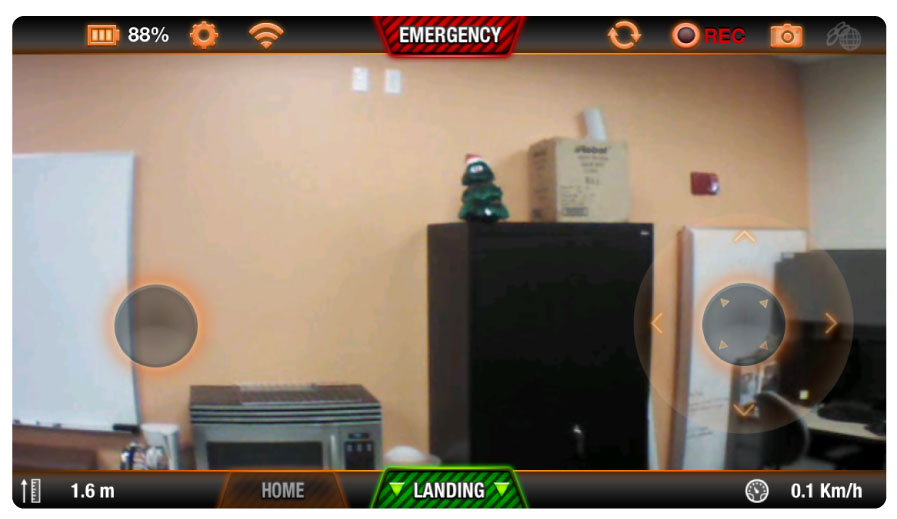}
	\caption{Piloting the quadcopter using AR.FreeFlight user interface}
	\label{fig:freeflight}
\end{figure} 

\subsection{Study 3: Visual Dataset Collection}
For this study, users were asked to collect an image dataset using our interface vs.\ capturing images using the AR.FreeFlight piloting application.  The Parrot developer community has created versions of the AR.FreeFlight user interface for iOS, Android, and Windows platforms; it is the official form of software control for the AR.Drone quadcopter. The piloting section of the AR.FreeFlight UI has a screen that shows the frontal and downward cameras, along with takeoff/land buttons, photo/video capture buttons, and two joysticks (Figure~\ref{fig:freeflight}).  Moving the quadcopter horizontally can be done through using the left joystick or tilting the tablet/phone.  The autonomous image capturing option offered by our system frees the participants from doing it manually. This option along with the target tracking feature ease the operation of image capturing.  Figure~\ref{fig:q3} shows that users were able to rapidly collect more images using our interface (significantly more according to a paired single tailed t-test at the $p<0.05$ level).  From Figure~\ref{fig:s3}, we can see that when participants used our interface they were able to acquire more diverse image views.  This variety in image characteristics is particularly valuable for training machine learning classifiers. 

After the experiments, we administered a post-questionnaire.   Key questions included:
\begin{compactenum}
\item Would you like to have an assistant agent helping you out with capturing images of the selected object automatically?  
\item Would you like to have an assistant agent helping you out with navigation while capturing the images? 
\item Would you use the proposed user interface to collect images for your own project? 
\end{compactenum}
The majority of the participants responded positively to all these questions, indicating a high level of satisfaction with the concept of the intelligent user interface.

\section{Conclusion}  

In this paper we introduced a smart user interface (SUI) that uses sketch-based control to facilitate drone navigation and visual dataset collection tasks.  Our implementation is platform-independent and can be accessed from any mobile device without prior installation.  Our experiments demonstrate that our interface outperforms standard commercial solutions, such as joystick and AR.Free Flight.   A key contribution is the use of adaptive correlation filters for visual tracking of objects in the semi-autonomous target selection mode.  The MOSSE filter is robust against many appearance changes and capable of executing at high frame rates.  We tested our platform in three different scenarios with participants from different age groups. The participants were able to robustly execute navigation patterns and collect visual datasets without crashing and expressed satisfaction with the user experience.  

\section{Acknowledgments}
Support for Saif Alabachi's study was provided by University of Technology (Baghdad). This work was partially funded by a UCF Reach for the Stars award.
 
\bibliography{bib}  
\bibliographystyle{aaai}

\end{document}